# Structural Phase Transitions in SrRh$_2$As$_2$

V. Zinth[1], V. Petricek[2], M. Dusek[2], and D. Johrendt*[1]

[1] *Department of Chemistry, Ludwig-Maximilians-Universität, 81377 München, Germany*

[2] *Institute of Physics, Cukrovarnicka 10, 16253 Praha, Czech Republic*

SrRh$_2$As$_2$ exhibits structural phase transitions reminiscent to those of BaFe$_2$As$_2$, but crystallizes with three polymorphs derived from the tetragonal ThCr$_2$Si$_2$-type structure. The structure of α-SrRh$_2$As$_2$ is monoclinic with $a$ = 421.2(1) pm, $b$ = 1105.6(2) pm, $c$ = 843.0(1) pm and β = 95° and was refined as a partially pseudo meroedric twin in the space group $P2_1/c$ with R1 = 0.0928. β-SrRh$_2$As$_2$ crystallizes with a modulated structure in the (3+1) dimensional superspace group $Fmmm$(10γ)σ00 with the unit cell parameters $a$ = 1114.4(3) pm, $b$ = 574.4(2) pm and $c$ = 611.5(2) pm and an incommensurable modulation vector **q** = (1, 0, 0.3311(4)). High temperature single crystal diffraction experiments confirm the tetragonal ThCr$_2$Si$_2$-type structure for γ-SrRh$_2$As$_2$ above 350°C. Electronic band structure calculations indicate that the structural distortion in α-SrRh$_2$As$_2$ is caused by strong Rh-Rh bonding interactions and has no magnetic origin as suggested for isotypic BaFe$_2$As$_2$.

## I. Introduction

The coupling of electronic and lattice degrees of freedom creates some of the most intriguing phenomena in solid state materials. Two well known manifestations of electron-lattice coupling are charge-/spin-density-waves (CDW, SDW)[1] and conventional superconductivity.[2] In superconductors, the coupling is usually weak, but it determines the critical temperature $T_c$. Stronger coupling increases $T_c$ to a certain limit, while too strong interactions can drive the system to a CDW state, where a structural distortion reduces the electronic energy.

Structural phase transitions associated to CDW instabilities have been observed in many metallic materials with quasi low-dimensional crystal structures, among them transition-metal chalcogenides[3,4] and oxides.[5] CDW ordering has often been considered as a manifestaion of the Peierls-instability, thus relying on nesting, which means that a piece of the Fermi surface can be translated by a vector **q** and superimposed on another piece of the surface. Meanwhile, reasonable doubts arise whether nesting is the only driving force of the CDW, and in fact, **q**-vectors extracted from Fermi surface nesting are in some cases inconsistent with experimental wave vectors,[6] even in the archetypical CDW compound TaSe$_2$.[7]

Layered crystal structures also constitute the basis of the high-temperature superconductors, both the copper oxides[8] and the iron arsenides.[9,10] The latter are build up by layers of edge-sharing

FeAs$_4$-tetrahedra, separated either by oxide layers as in LaOFeAs with ZrCuSiAs-type structure[11] or by alkaline earth metals as in BaFe$_2$As$_2$ with ThCr$_2$Si$_2$-type structure.[12] These non-superconducting parent compounds exhibit SDW-type phase transitions, accompanied by a reduction of the space group symmetry from tetragonal to orthorhombic.[13,14] In these materials, the SDW wave vectors perfectly coincide with the Fermi surface nesting, which is also believed to play a certain role in the pairing mechanism.[15]

The orthorhombic low-temperature structure of BaFe$_2$As$_2$ was first classified[16] as isotypic to the β-SrRh$_2$As$_2$-type in the space group *Fmmm*.[17] Indeed also rhodium compound transforms to the tetragonal ThCr$_2$Si$_2$-type (γ-SrRh$_2$As$_2$, space group *I4/mmm*), but at much higher temperature (555 K) in comparison with BaFe$_2$As$_2$ (140 K)[14] and SrFe$_2$As$_2$ (210 K).[18] Furthermore, the ortho-rhombic lattice distortion, measured as $\delta = |a-b|/(a+b)$, is about one order of magnitude larger in SrRh$_2$As$_2$ ($\delta \approx 10^{-2}$) than in BaFe$_2$As$_2$ ($\delta \approx 10^{-3}$). Despite the striking symmetry congruence of the *I4/mmm* ↔ *Fmmm* structural transitions in SrRh$_2$As$_2$ and BaFe$_2$As$_2$, both are not expexted to have the same origin. Since SrRh$_2$As$_2$ carries no magnetic moment, the driving force is not magneto-elastic coupling by antiferromagnetic SDW ordering as in BaFe$_2$As$_2$. Thus, the β → γ transition of SrRh$_2$As$_2$ may be assigned to a possible CDW instability of the RhAs$_{4/4}$ layers. However, recently published band structure calculations[19] revealed no hint to an instability, however, the electronic susceptibilty has not yet been analyzed in detail. Furthermore, β-SrRh$_2$As$_2$ transforms to a third (α-) modification, which is stable below 555 K. The structure of α-SrRh$_2$As$_2$ was assumed to be closely related to the orthorhombic BaNi$_2$Si$_2$-type (space group *Cmcm*), but could not be successfully refined.

In order to shed light on several open issues regarding the crystal structures and phase transitions of polymorphic SrRh$_2$As$_2$ and its relationships to those occurring in the isostructural iron arsenides, we have synthesized SrRh$_2$As$_2$ and conducted detailed single crystal X-ray experiments.

## II. Experimental Details

Powder samples of α-SrRh$_2$As$_2$ were synthesized by heating stoichiometric mixtures of the elements in alumina crucibles that were sealed in silica tubes under an atmosphere of purified argon. The mixtures were heated to 620 °C, kept at this temperature for 10 h and cooled down to room temperature. The reaction products were homogenized and annealed at 1000 °C for 30 hours several times. X-ray powder patterns have been measured on a STOE Stadi-P diffractometer (Cu-K$_{\alpha1}$ radiation). The TOPAS[20] package was used for Rietveld refinements. The powder diffraction pattern of α-SrRh$_2$As$_2$ together with the Rietveld refinement are shown in Fig. 1.

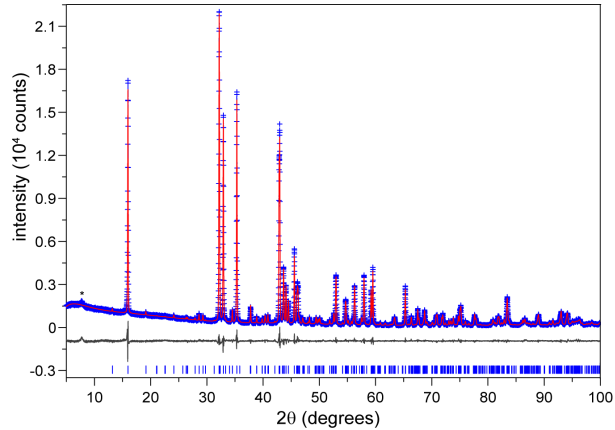

FIG. 1. X-ray powder pattern and Rietveld refinement of α-SrRh$_2$As$_2$

Single crystals of α-SrRh$_2$As$_2$ were grown in a Pb/Bi-flux (55 wt % Bi, ten times surplus of flux) by heating to 1100 °C, holding the temperature for 30 h, cooling to 160 °C with 2 °C/h and then quenching the sample. Single crystals of β-SrRh$_2$As$_2$ were obtained in a similar way using 45 wt % Bi and a cooling rate of 30 °C/h. The flux was dissolved in HAc/H$_2$O$_2$, and revealed platelike crystals with a tendency to cleave. The crystals were checked by Laue photographs using white molybdenum radiation. Single crystal intensity data of α- and γ-SrRh$_2$As$_2$ was recorded on a STOE IPDS imaging plate detector (Ag-K$_α$, graphite monochromator, φ-scan) equipped with a Heatstream (Stoe & Cie Gmbh, Darmstadt, Germany) device for high temperature measurements. Single crystal intensity data of β-SrRh$_2$As$_2$ was collected on a Gemini four-circle diffractometer equipped with an Atlas CCD detector. For structure solution and refinement the SHELX suite of programs[21-22] (α-SrRh$_2$As$_2$ and γ-SrRh$_2$As$_2$) and the program Jana2006[23] (β-SrRh$_2$As$_2$) were used. The electronic structure and Crystal Orbital Hamilton Function (COHP) of α-SrRh$_2$As$_2$ and γ-SrRh$_2$As$_2$ were calculated from self-consistent TB-LMTO-ASA[24] potentials and wave functions[25] using density-functional (DFT) methods

## III. Results

### A Crystal structures

#### 1. α-SrRh$_2$As$_2$

Diffraction patterns of the room temperature phase pretended orthorhombic *mmm* symmetry as reported in Ref.[17], but all attempts at finding a satisfactory structure model failed. A careful inspection of the pattern revealed multiple twinning, but only two main domains contribute considerably as shown in Figure 2a. Their unit cell dimensions are *a* = 421.2(1) pm, *b*

= 1105.6(2) pm, $c$ = 843.0(1) pm and $\beta$ = 95 °, respectively. The latter is supported by observations in single crystals of $\alpha$-SrRh$_2$As$_2$ with about 10 % of Sr substituted by Ba, where an easier kind of twinning occurs and the unit cell dimensions can easily be identified. The domains shown in figure 2a transform onto each other by a twofold rotation around [1 0 2] that feigns the two times larger pseudo unit cell $a \approx$ 842 pm, $b \approx$ 1106 pm, $c \approx$ = 842 pm and $\beta$ = 95 °. or a four times larger pseudo-orthorhombic unit cell with $A = -2a-c$ = 1138 pm, $B = -b$ = 1105.6 pm, $C = -2a+c$ = 1243.2 pm, $\alpha$ = 90 °, $\beta$ = 90.04 °, $\gamma$ = 90 °. Integration of the intensity data was performed using this doubled monoclinic pseudo unit cell. An absorption correction was performed using the shape of the crystal as obtained by the diffractometer's video system. Then the reflections were transformed according to the two domains, stored into a .hklf5 file and merged with the program mergehklf5[26]. The structure was solved and refined in the space group $P2_1/c$ with R1 = 0.0928 (Table 1).

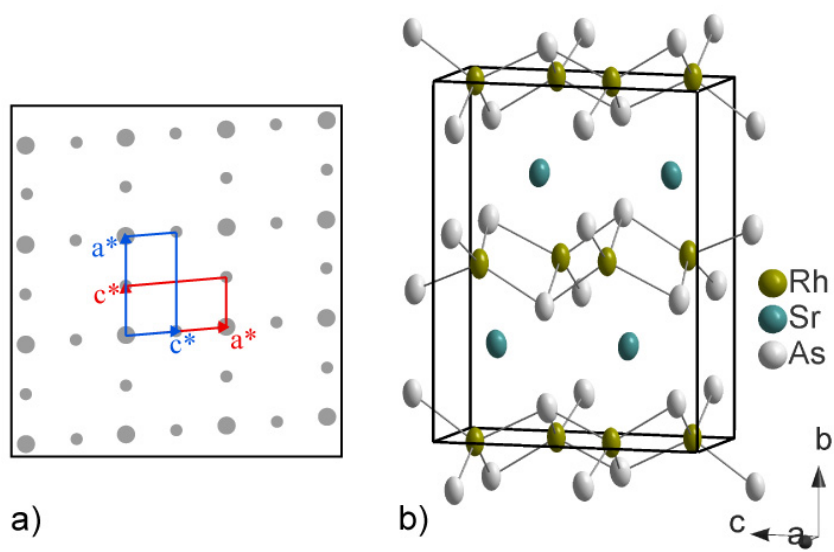

FIG. 2. a) Schematical picture of the reciprocal space of $\alpha$-SrRh$_2$As$_2$ with the unit cells of the two twin domains, b) unit cell of $\alpha$-SrRh$_2$As$_2$, the anisotropic atomic displacement parameters (95 % probability) are shown

Selected interatomic distances and bond angles of $\alpha$-SrRh$_2$As$_2$ are compiled in Table 2. The shortest As-As distances between the RhAs$_4$-tetrahedron layers is 313.0 pm, thus no significant interlayer bonding is expected. The Rh-As distances range from 238.8 to 251.1 pm and the As-Rh-As angles in the RhAs$_4$-tetrahedra from 79.24 ° to 123.89 °. The strong distortions of the RhAs$_4$ - tetrahedron layers are visible in Fig. 2b, which shows the unit cell of $\alpha$-SrRh$_2$As$_2$. The distortion is best be grasped by looking at the positions of the Rh atoms within the RhAs-layers (Fig. 3). In contrast to the regular square network of Rh in the tetragonal high temperature phase with a constant distance of 290 pm between the Rh atoms, Rh-Rh distances in $\alpha$-SrRh$_2$As$_2$ vary from 284.7 pm to 378.5 pm, and chains of four Rh atoms with short distances alternating with one longer distance between the chains.

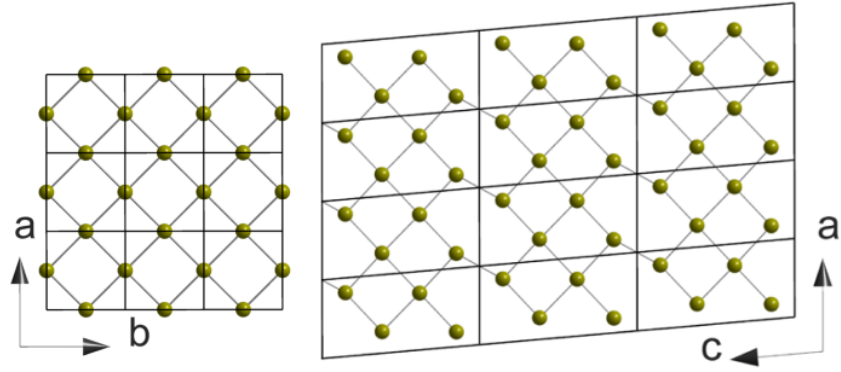

FIG. 3. Comparison of the rhodium networks in γ- (left) and α-SrRh$_2$As$_2$ (right)

Table 1 crystallographic data of α-SrRh$_2$As$_2$

| | | | | |
|---|---|---|---|---|
| Empirical formula | | SrRh$_2$As$_2$ | | |
| Crystal system, space group | | Monoclinic, $P2_1/c$, No. 14 | | |
| $a, b, c$ (pm), $\beta$ (°) | | 421.2(1), 1105.6(2), 843.0(1), 95.06(2) | | |
| Cell volume (nm$^3$) | | 0.3920(1) | | |
| Molar mass (g/mol) | | 443.3 | | |
| Calculated density (g/cm$^3$), $Z$ | | 7.53, 4 | | |
| Radiation type, $\lambda$ (Å) | | Ag-K$_\alpha$, 0.56087 | | |
| $2\theta$ range | | 4.6 - 60.9 | | |
| Transmission (min, max) | | 0.1960, 0.5600 | | |
| Absorption coefficient (mm$^{-1}$) | | 20.46 | | |
| Total number of reflections | | 12430 | | |
| Independent reflections, $R_{meas}$ | | 3652, 0.2188 | | |
| Reflections with $I>2\sigma(I)$, $R_\sigma$ | | 1582, 0.1545 | | |
| Refined parameters, Goodness-of-Fit on $F^2$ | | 47, 0.86 | | |
| $R1, wR2$ ($I>2\sigma(I)$) | | 0.0928, 0.2312 | | |
| $R1, wR2$ (all data) | | 0.1891, 0.3107 | | |
| Largest residual peak, hole $e^-/\text{Å}^3$ | | 5.654, -6.042 | | |
| Twin fraction | | 51.7(3) % | | |
| Atomic parameters | | | | |
| | | x | y | z | U$_{eq}$ |
| Sr | 4e | 0.2478(4) | 0.7539(1) | 0.1341(2) | 0.0184(3) |
| Rh1 | 4e | 0.2245(3) | 0.5067(2) | 0.3807(2) | 0.0166(3) |
| Rh2 | 4e | 0.3116(4) | 0.0093(2) | 0.3522(2) | 0.0179(3) |
| As1 | 4e | 0.1868(4) | 0.4072(2) | 0.1126(2) | 0.0185(3) |
| As2 | 4e | 0.2749(4) | 0.1229(2) | 0.1108(2) | 0.0171(3) |

Table 2 interatomic distances (pm) and angles in α-SrRh$_2$As$_2$

| Rh1 - | Distance | Rh2 - | Distance | Sr3 - | Distance |
|---|---|---|---|---|---|
| As4 | 240.8(2) | As4 | 238.8(2) | As4 | 317.0(2) |
| As4 | 246.4(2) | As5 | 239.1(2) | As4 | 319.4(2) |
| As4 | 247.0(2) | As5 | 242.4(2) | As5 | 319.5(2) |
| As5 | 251.1(2) | As5 | 248.3(2) | Rh2 | 327.8(2) |
| Rh2 | 286.4(3) | Rh2 | 284.7(3) | As4 | 330.1(2) |
| Rh1 | 288.4(2) | Rh1 | 286.4(3) | Rh2 | 337.1(2) |
| Rh2 | 289.0(2) | Rh1 | 288.8(2) | As5 | 338.8(2) |
| Rh1 | 294.3(3) | | | Rh1 | 340.1(2) |
| | | | | Rh1 | 342.8(2) |
| As5 - | | | | Rh1 | 344.4(2) |
| As5 | 313.0(1) | | | | |
| As4 | 317.0(1) | | | | |

| Angle | | | |
|---|---|---|---|
| As4-Rh1-As4 | 105.68(8) | As4-Rh2-As5 | 110.23(7) |
| As4-Rh1-As4 | 107.53(7) | As4-Rh2-As5 | 110.89(8) |
| As4-Rh1-As5 | 117.5(1) | As4-Rh2-As5 | 123.89(9) |
| As4-Rh1-As4 | 117.2(1) | As5-Rh2-As5 | 122.0(1) |
| As4-Rh1-As5 | 104.02(7) | As5-Rh2-As5 | 108.54(8) |
| As4-Rh1-As5 | 105.46(8) | As5-Rh2-As5 | 79.24(8) |

## 2. β-SrRh$_2$As$_2$

First measurements of β-SrRh$_2$As$_2$ single crystals revealed orthorhombic all face centered symmetry with unit cell parameters $a$ = 1114.4(3) pm, $b$ = 574.4(2) pm and $c$ = 611.5(2) pm in agreement with Ref.[17]. An initial refinement using the space group *Fmmm* converged to R1 = 0.055, but the anisotropic displacement parameters were not acceptable and clearly hint at a lower symmetry. The inspection of the diffraction pattern revealed additional satellite spots surrounding each Bragg position of the orthorhombic cell as shown in Fig. 4. These strong satellite reflections were indexed using the nearly commensurable modulation vector **q** = (1, 0, ≈1/3). Additional, very weak spots were detected at **q** = (0.5, 0, ≈1/6) and further, barely detectable reflections exist for some (h+1/2, k, l+1/2) (possibly 3$^{rd}$ order satellites) and (h, k+1/2, l+1/2). Using the (0.5, 0, ≈1/6) reflections as first order satellites, we have the rather unusual case that the second order satellites are much stronger than the first order ones. This clearly indicates that the second order components of the modulation functions are predominant. As a first approach, we have neglected the weak first order spots and tried to find a solution using **q** = (1, 0, 0.3311(4)) as the modulation vector, which is compatible with the (3+1)-dimensional superspace group *Fmmm*(10γ)σ00.

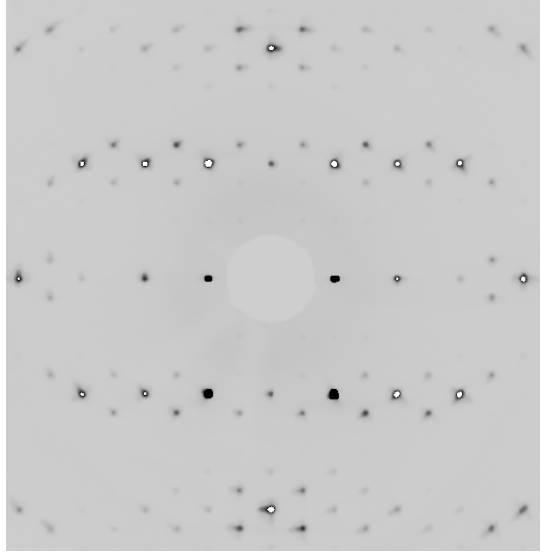

FIG. 4. h0l-layer of β-SrRh$_2$As$_2$ with strong satellite spots at **q** = (1, 0, 0.3311(4)).

A comparison of the lattice parameters shows similarities between the above mentioned pseudo orthorhombic lattice $A$, $B$, $C$ of α-SrRh$_2$As$_2$ and the cell parameters of β-SrRh$_2$As$_2$ ($a \approx A$, $b \approx B/2$, $c \approx C/2$). Thus it seems possible that the α-phase is still partially present in the crystal of the β-phase. This would cause additional reflections with ($h$, $k+1/2$, $l+1/2$) and violations of the $F$-centering, both of which we observe. A multiphase refinement of the crystal gives a significantly better fit for the collected data and yields a volume fraction of 68 % β-phase and 16 % for each of the α-phase twin components. The refinement data are summarized in Tables 3, 4 and 5. The **q**-vector is close to commensurable, but the best refinement results were obtained with the incommensurate model. The overlap option for closest reflections implemented in Jana2006 was used. Both first and second order satellite reflections are reasonably well refined (see Table. 3).

The strongest modulation concerns the position of the Rh-atom in $c$-direction (Fig. 5a). A crenel function was used to describe the modulation, combined with a positional modulation function expressed by Legendre polynomial. No modulation of the Rh atoms in $a$- and $b$- directions were observed. The strong modulation of the Rh atom influences the adjoining As atom, which shows a positional modulation in $c$- direction that can be described very well by a combination of one sinus and one cosinus term (Fig 5b). For the As atom also a second positional modulation with smaller amplitude is observed in $a$- direction (Fig 6a). The Sr atoms show pretty much the same modulation in $a$-direction (Fig. 6b). The modulation function of the ADP parameters was also refined, with the biggest changes found for the ADP parameters of the As atom (see Table 4).

Table 3 crystallographic data of β-SrRh$_2$As$_2$

| | |
|---|---|
| Empirical formula | SrRh$_2$As$_2$ |
| Crystal system, space group | orthorhombic, Fmmm(10γ)σ00 |
| $a, b, c$ (pm) | 1114.4(3), 574.4(2), 611.5(2) |
| Cell volume (nm$^3$) | 0.3914 (2) |
| Molar mass (g/mol) | 443.3 |
| Calculated density (g/cm$^3$), $Z$ | 7.52, 4 |
| Radiation type, $\lambda$ (Å) | Mo-K$_\alpha$, 0.7107 |
| $2\theta$ range | 5.7272 - 58.6562 |
| Transmission (min, max) | 0.23117, 1.000 |
| Absorption coefficient (mm$^{-1}$) | 38.425 |
| Total number of reflections | 23001 |
| Independent reflections, R$_{int}$ | 1271, 0.0470 |
| Reflections with $I>3\sigma(I)$, R$_\sigma$ | 833, 0.0105 |
| Main reflections: I>3σ(I)/ all | 327/ 479 |
| 1. order satellites: I>3σ(I)/ all | 225/ 258 |
| 2. order satellites: I>3σ(I)/all | 281/ 534 |
| Refined parameters, GooF | 47 , 5.90 |
| $R$(obs)($I>3\sigma(I)$)/R(obs)(all) | 0.0563/ 0.0918 |
| Main reflections | 0.0424/ 0.0567 |
| 1. order satellites | 0.0569/ 0.0658 |
| 2. order satellites | 0.0911/ 0.1952 |
| $w$R($I>3\sigma(I)$)/$w$R(all) | 0.1089/ 0.1260 |
| Main reflections | 0.0982/ 0.0991 |
| 1. order satellites | 0.1045/ 0.1054 |
| 2. order satellites | 0.1364/ 0.1942 |
| Residual peak, hole $e^-$/Å$^3$ | 6.67/-5.66 |

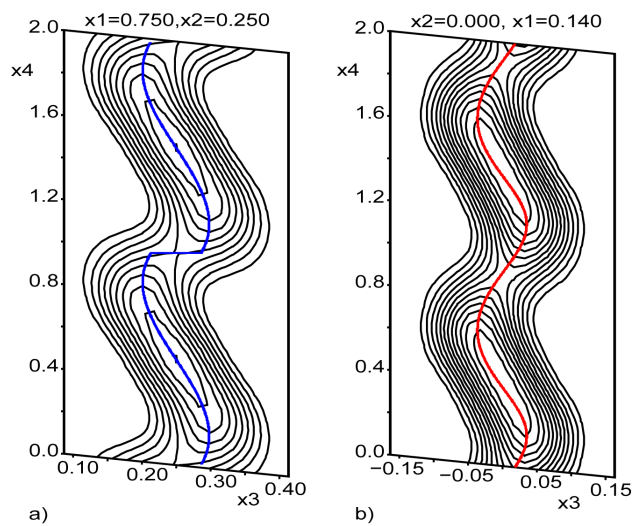

FIG. 5. Fourirer map with a) positional modulation of Rh in $c$-direction, b) of As in $c$-direction

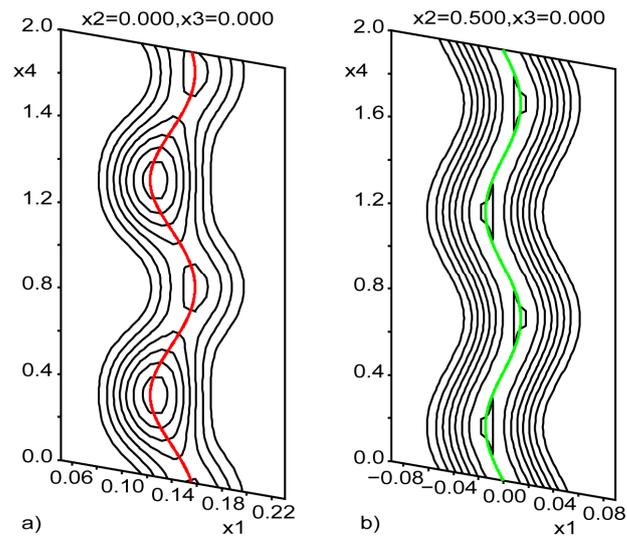

FIG. 6. Fourier map with a) Positional modulation of As in *a*-direction, b) of Sr in *a*-direction.

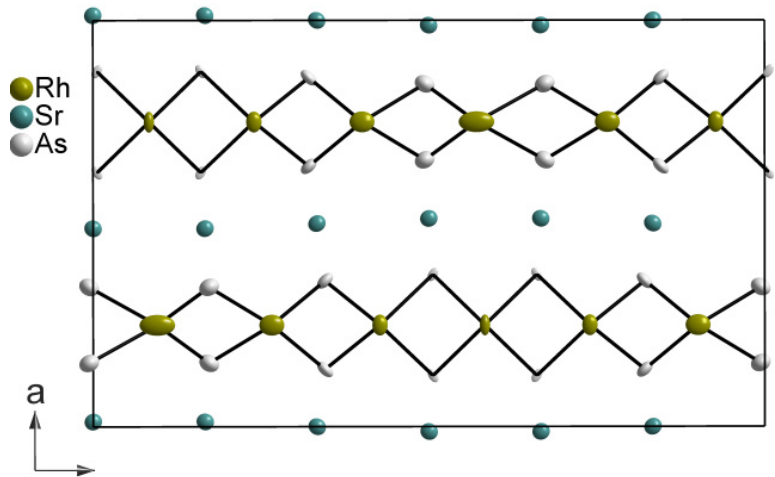

FIG. 7. Modulated structure of β-SrRh$_2$As$_2$, projection along *b*.

Table 4 parameters of the modulation functions of β-SrRh$_2$As$_2$

| Atom | x, y, z | $U_{11}$ | $U_{22}$ | $U_{33}$ |
|---|---|---|---|---|
| Rh, 8d | 0.75, 0.25, 0.25 | 0.0252(7) | 0.0083(5) | 0.0282(7) |
| Sr 4d | 0, 0.5, 0 | 0.0163(11) | 0.0103(12) | 0.0174(10) |
| As 8d | 0.1410(2), 0, 0 | 0.0198(8) | 0.0312(10) | 0.0204(7) |

Fourier coefficients of the modulation functions

| | | $u_x^S$ | $u_y^S$ | $u_z^S$ | $u_x^C$ | $u_y^C$ | $u_z^C$ |
|---|---|---|---|---|---|---|---|
| Rh | 1. order | | | | | | |
| | | 0 | 0 | −0.0621(5) | 0 | 0 | 0 |
| | 2. order | | | | | | |
| | | 0 | 0 | 0.0255(6) | 0 | 0 | 0 |
| Sr | | −0.0141(2) | 0 | 0 | 0 | 0 | 0 |
| As | | −0.01136(14) | 0 | 0.0278(3) | 0.01391(17) | 0 | 0.0227(2) |

| Crenel function | width | centre |
|---|---|---|
| Rh | 1 | 0.5 |

Coefficients of the modulation function of the ADP parameters

| Atom | term | $U_{11}$ | $U_{22}$ | $U_{33}$ |
|---|---|---|---|---|
| Rh | s | 0 | 0 | 0 |
| | c | -0.0000(9) | 0.0034(8) | 0.00442(15) |
| Sr | s | 0 | 0 | 0 |
| | c | 0 | 0 | 0 |
| As | s | -0.0040(9) | -0.0220(6) | -0.0081(4) |
| | c | 0.0049(12) | 0.0269(7) | 0.0099(5) |
| | | $U_{12}$ | $U_{13}$ | $U_{23}$ |
| Rh | s | 0.0013(6) | 0 | 0 |
| | c | 0 | 0 | 0 |
| Sr | s | 0 | 0 | 0 |
| | c | 0 | -0.0015(6) | 0 |
| As | s | 0 | -0.0079(7) | 0 |
| | c | 0 | -0.0064(5) | 0 |

Table 5 crystallographic data of α-SrRh$_2$As$_2$ in the multiphase refinement

| Atom | x | y | Z | $U_{iso}$ |
|---|---|---|---|---|
| Sr1 | 0.246(2) | 0.757(2) | 0.1294(11) | 0.020(3) |
| Rh1 | 0.215(3) | 0.5007(13) | 0.3911(15) | 0.0104(19) |
| Rh2 | 0.308(5) | 0.0046(14) | 0.348(2) | 0.028(2) |
| As1 | 0.207(4) | 0.4015(15) | 0.128(2) | 0.038(5) |
| As2 | 0.273(3) | 0.1244(11) | 0.1169(16) | 0.021(3) |

twinning matrices: β-phase to α-phase

1. (0 1 0 -0.5 0 -1 -0.5 0 1)    2. (0 -1 0 -0.5 0 -1 0.5 0 -1)

Figure 7 shows the modulated structure of β-SrRh$_2$As$_2$ (view along the *b*-axis, three unit cells in *c*-direction are shown). Both the slight modulations of As and Sr in *c*-direction and the modulations of the ADP parameters are visible. Fig. 8c illustrates the modulation within the RhAs$_4$ tetrahedra planes (only the Rh atoms are shown). The modulation leads to a strong elongation of every seventh Rh-Rh distance to 358.0 pm along *c*, while the adjoining distances have a length of 313.2 pm and 295.8 pm and the shorter ones as well as all distances parallel to *b* remain close to 287 pm. Fig. 8b shows the *t*-plot of the Rh-Rh distances. As expected, the biggest change of the Rh-Rh distances takes place near the "jump" of the Rh modulation function. The Rh-As distances (*t*-plot shown in Fig. 8a) vary from ~227 pm to 257 pm, the variation is bigger than in α-SrRh$_2$As$_2$. Both the minimal and maximal distances occur near the jump of the modulation function of the Rh-atom, as the As atoms adjust to their bonding partner. Near this point, also enlarged atomic displacement parameters are observed (Fig. 7).

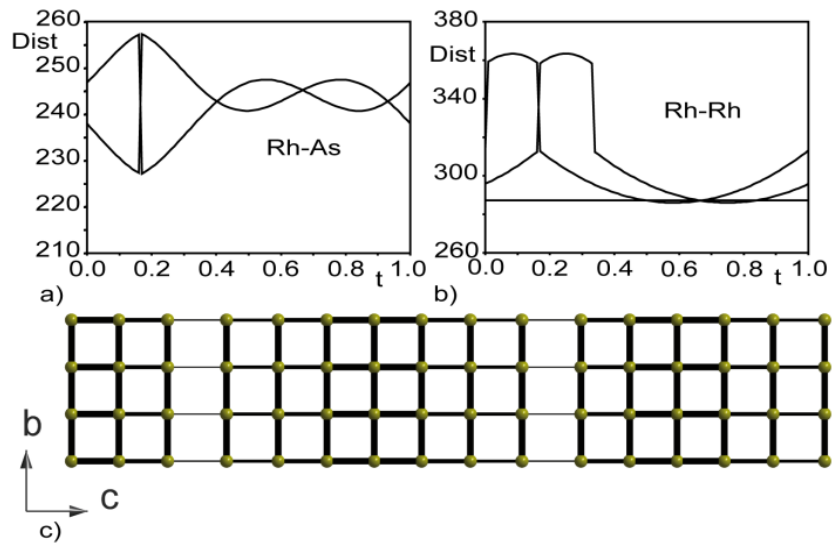

FIG. 8. a) *t*-plot of the Rh-As distances (pm), b) *t*-plot of the Rh-Rh distances (pm), c) Rh-Rh network in β-SrRh$_2$As$_2$, elongated distances are shown with thinner lines

This model describes the essential characteristics of the modulated structure of β-SrRh$_2$As$_2$, even though the much weaker satellite reflections with **q** = (0.5, 0, ≈1/6) are still neglected. These will probably allow to describe the modulation in more detail, and maybe help to improve both displacement parameters and Rh-As distances close to the jump of the modulation. However, due to the very weak intensity of these spots, we were not able to measure them with the accuracy required to improve our model.

Previous publications[17] mentioned the β-phase as a high temperature polymorph stable between about 190 and 282 °C. We confirm these results as we could observe the phase transition from α- to β- phase with high temperature powder diffraction. A flux synthesized crystal of β-SrRh$_2$As$_2$ showed still the same satellite reflections found at room temperature at 290 °C, but the quality was not sufficient for further x-ray analysis. Upon cooling, the crystal transformed to α-SrRh$_2$As$_2$. The α–phase is the stable polymorph at room temperature, which is nicely supported by the fact that β-SrRh$_2$As$_2$ easily transforms to α-SrRh$_2$As$_2$ at room temperature under pressure of approximately 1-2 GPa in a hand press (Fig. 9). The fraction of α-phase in the single crystal of β-SrRh$_2$As$_2$ used for structure determination increased from 7 % to about 30 % within one year at ambient conditions. Magnetic susceptibility measurements from 300 to 1.8 K on a sample of β-SrRh$_2$As$_2$ with some fraction of α-SrRh$_2$As$_2$ showed values typical for a Pauli-paramagnetic metal, although some traces of ferromagnetic impurity were observed.

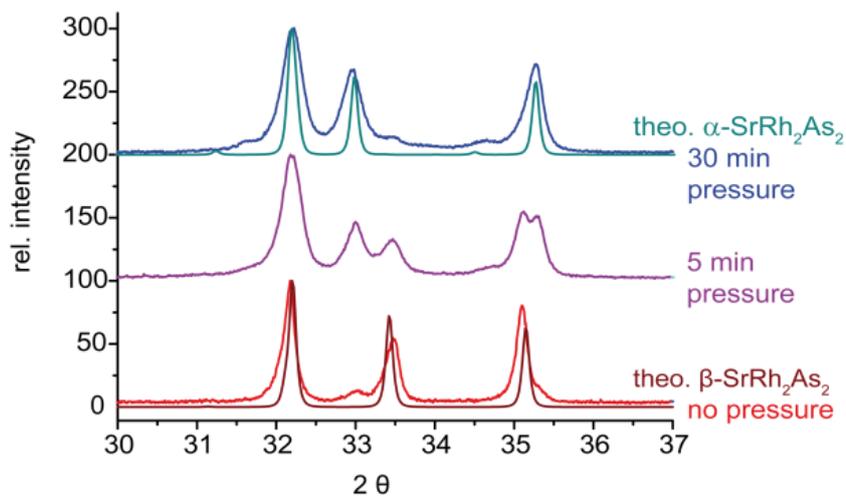

FIG. 9. β-SrRh$_2$As$_2$ transforms to α-SrRh$_2$As$_2$ under pressure

### 3. γ-SrRh$_2$As$_2$

Several attempts to obtain powder samples or single crystals of γ-SrRh$_2$As$_2$ by quenching from high temperatures remained unsuccessful, but the diffraction pattern of a β-SrRh$_2$As$_2$ crystal measured at 350 °C showed tetragonal symmetry. As described in the literature[17] γ-SrRh$_2$As$_2$ crystallizes in the tetragonal ThCr$_2$Si$_2$-type structure (Table 6). The lattice parameter $c$ is 1156.1(3) pm at 350 °C, thus elongated compared to 1143.1(6) pm as reported for the γ-phase at room temperature.[17] Also the ADP parameters are enlarged due to the higher temperatures.

Table 6. Crystallographic data of γ-SrRh$_2$As$_2$ at 350 °C

| Empirical formula | SrRh$_2$As$_2$ |
|---|---|
| Crystal system, space group | Tetragonal, *I4/mmm*, No. 139 |
| *a, c* (pm) | 412.68(6), 1156.1(3) |
| Cell volume (nm$^3$) | 0.19688(6) |
| Molar mass (g/mol) | 443.3 |
| Calculated density (g/cm$^3$), Z | 7.477, 2 |
| Radiation type, λ (Å) | Ag-K$_\alpha$, 0.56087 |
| Temperature (K) | 623 |
| 2θ range | 4.6 - 60.9 |
| Transmission (min, max) | 0.0605, 0.6104 |
| Absorption coefficient (mm$^{-1}$) | 20.32 |
| Total number of reflections | 2268 |
| Independent reflections, R$_{int}$ | 213, 0.1229 |
| Reflections with *I*>2σ(*I*),R$_\sigma$ | 168, 0.0604 |
| Refined parameters, Goodness-of-Fit on *F*$^2$ | 8, 1.158 |
| *R*1,*wR*2 (*I*>2σ(*I*)) | 0.0376, 0.1392 |
| *R*1,*wR*2 (all data) | 0.0516, 0.1623 |
| Largest residual peak, hole *e*$^-$/Å$^3$ | 4.164, -1.494 |
| Atomic parameters | |
| Sr   2a   (0,0,0) | U$_{eq}$ = 0.0211(4) |
| Rh   4d   (0,0.5,0.25) | U$_{eq}$ = 0.0286(4) |
| As   4e   (0,0,*z*) *z* = 0.3637(2) | U$_{eq}$ = 0.0255(4) |

**B Electronic structure and discussion**

The distortions of the ThCr$_2$Si$_2$-type structure in SrRh$_2$As$_2$ mainly affect the Rh-Rh bonds within the RhAs$_4$-layer, while the changes of other bond lengths may be regarded as consequences of that. It seems probable that the origin of the polymorphism lies in the Rh-Rh bonding situation, which has been analyzed by the Crystal Orbital Hamilton Population (COHP) method. Fig. 10 shows COHP curves of the Rh-Rh bonds in tetragonal γ-SrRh$_2$As$_2$ with ThCr$_2$Si$_2$-type structure and monoclinic α-SrRh$_2$As$_2$. For γ-SrRh$_2$As$_2$ (Fig. 10a), the calculations show strongly Rh-Rh antibonding states around the Fermi energy E$_F$, caused by *dd*σ*-type interactions of the in-plane rhodium orbitals within the square network as shown in Fig. 2 (right). As those interactions are quite unfavourable, one would expect the situation to be improved in the low temperature phases. The very similar situation of SrRh$_2$P$_2$ has been analyzed earlier in detail.[20] If we look at the structural changes within this Rh-Rh network for the low temperature phases, we first find the distortion from tetragonal to orthorhombic in β-SrRh$_2$As$_2$. The additional modulation causes strong elongations of several Rh-Rh-bonds (Fig. 8c). Every seventh bond along *c* is stretched to

358.0 pm, with the adjoining bonds elongated slightly to 313.2 and 295.8 pm. All other bonds in this direction and all bonds along *b* are almost the same as in the γ-phase, with 287.12 pm and 287.4 pm compared to 290 pm in γ-SrRh$_2$As$_2$. This elongation of several bonds without much shortening of others is also reflected by the lattice parameters: The identical diagonals in the tetragonal γ-phase are $a*\sqrt{2}$ = 581.5 pm, and split into 574.4 pm (−7 pm) and 611.5 pm (+30 pm) in the β-phase. The monoclinic distortion in α-SrRh$_2$As$_2$ is even more successful in elongating Rh-Rh distances (Fig. 2): every fourth distance is widened to 378 pm. The effect of the structural changes on the bonding situation in α-SrRh$_2$As$_2$ can be studied in the Rh-Rh COHP that is shown in Fig. 10 (right). Most of the antibonding states that were present at E$_F$ in γ-SrRh$_2$As$_2$ have been lowered in energy, and are thus less unfavourable. This can be considered as a typical CDW scenario, where compounds compensate highly symmetric but unfavourable bonding situations by splitting bonds in longer and shorter ones yielding lower symmetry structures. More detailed calculations and analysis of the electronic structure including the Fermi surface topology are still necessary to obtain details of a CDW scenario, but are beyond the scope of this article.

In spite of some similarities, this situation is clearly different from that in BaFe$_2$As$_2$, where the orthorhombic distortion is much smaller and intimately connected to magnetic interactions. Our results suggest that the much more pronounced effect in SrRh$_2$As$_2$ is caused by the larger overlap the in-plane 4*d*-orbitals responsible for strong Rh-Rh bonding. The higher electron count of SrRh$_2$As$_2$ places the Fermi-level just in the middle of the Rh-Rh antibonding *ddσ\**-bands which can be stabilized by splitting in a structure with lower symmetry. This is different in BaFe$_2$As$_2$, where the Femi-level cuts Fe-Fe antibonding bands of *ddπ\** character that overlap too weak to cause a strong lattice distortion.

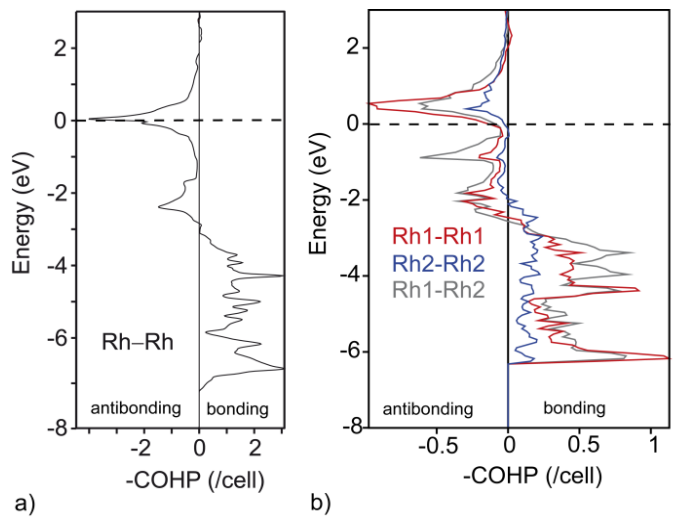

FIG. 10. a) Rh-Rh-COHP of tetragonal γ-SrRh$_2$As$_2$ b) Rh-Rh-COHP of monoclinic α-SrRh$_2$As$_2$

## IV Conclusions

In summary, we have synthesized single crystals and powder samples of SrRh$_2$As$_2$ and studied the structures of three polymorphs in detail. We solved the up to now unknown structure of α-SrRh$_2$As$_2$ that crystallizes in the monoclinic space group $P2_1/c$ with $a$ = 421.2(1) pm, $b$ = 1105.6(2) pm, $c$ = 843.0(1) pm and $β$ = 95 ° and is twinned. We revealed that the previous structural model for β-SrRh$_2$As$_2$ only describes the average structure. A structural modulation with **q** = (0.5, 0, 0.1655) was found, where second order satellites are much stronger than first order satellites. Taking into account the strong satellites we present an incommensurate structure in the (3+1) dimensional superspace group $Fmmm(10γ)σ00$ with the unit cell parameters $a$ = 1114.4(3) pm, $b$ = 574.4(2) pm and $c$ = 611.5(2) pm and a modulation vector **q** = (1, 0, 0.3311(4)). The strong modulation of the Rh atom could be described by a crenel function and leads to a variation of the Rh-Rh distances along $c$ from to 287 pm to 358 pm. For the γ-phase, high temperature single crystal data confirm the ThCr$_2$Si$_2$-type structure of γ-SrRh$_2$As$_2$ as reported in literature. Our DFT calculations with COHP bonding analysis show that distortion and elongation of Rh-Rh bonds lead to lowering in energy of antibonding states in α-SrRh$_2$As$_2$ compared to tetragonal γ-SrRh$_2$As$_2$, thus the driving force of the lattice distortions comes from Rh-Rh bonding and has no magnetic origin as suggested for BaFe$_2$As$_2$.

## Acknowledgement

This work was financially supported by the DFG within the priority program SPP 1458 "High-Tc superconductivity in iron pnictides" under project JO257/6-1.

## Notes and references